\documentclass[aps,prb,twocolumn,showpacs,preprintnumbers,superscriptaddress]{revtex4}
\usepackage{graphicx}
\usepackage{dcolumn}

\newcommand{\etal}{{\it et al.}}
\begin{document}

\title{Evidence of Strong Correlations and Coherence-Incoherence Crossover in the Iron Pnictide Superconductor KFe$_{2}$As$_{2}$}
\author{F. Hardy}
\email[]{Frederic.Hardy@kit.edu}
\affiliation{Karlsruher Institut f\"ur Technologie, Institut f\"ur Festk\"orperphysik, 76021 Karlsruhe, Germany}
\author{A. E. B\"ohmer}
\affiliation{Karlsruher Institut f\"ur Technologie, Institut f\"ur Festk\"orperphysik, 76021 Karlsruhe, Germany}
\author{D. Aoki}
\affiliation{INAC/SPSMS, CEA Grenoble, 38054 Grenoble, France}
\affiliation{IMR, Tohoku University, Oarai, Ibaraki 311-1313, Japan}
\author{P. Burger}
\affiliation{Karlsruher Institut f\"ur Technologie, Institut f\"ur Festk\"orperphysik, 76021 Karlsruhe, Germany}
\author{T. Wolf}
\affiliation{Karlsruher Institut f\"ur Technologie, Institut f\"ur Festk\"orperphysik, 76021 Karlsruhe, Germany}
\author{P. Schweiss}
\affiliation{Karlsruher Institut f\"ur Technologie, Institut f\"ur Festk\"orperphysik, 76021 Karlsruhe, Germany}
\author{R. Heid}
\affiliation{Karlsruher Institut f\"ur Technologie, Institut f\"ur Festk\"orperphysik, 76021 Karlsruhe, Germany}
\author{P. Adelmann}
\affiliation{Karlsruher Institut f\"ur Technologie, Institut f\"ur Festk\"orperphysik, 76021 Karlsruhe, Germany}
\author{Y. X. Yao}
\affiliation{Ames Laboratory US-DOE, Ames, Iowa 50011, USA}
\author{G. Kotliar}
\affiliation{Department of Physics and Astronomy, Rutgers University, Piscataway, New Jersey 08854, USA}
\author{J. Schmalian}
\affiliation{Karlsruher Institut f\"ur Technologie, Institut f\"ur Theorie der Kondensierten Materie, 76128 Karlsruhe, Germany}
\author{C. Meingast}
\affiliation{Karlsruher Institut f\"ur Technologie, Institut f\"ur Festk\"orperphysik, 76021 Karlsruhe, Germany}

\date{\today}
\begin{abstract}
Using resistivity, heat-capacity, thermal-expansion, and susceptibility measurements we study the normal-state behavior of KFe$_{2}$As$_{2}$. We find that both the Sommerfeld coefficient ($\gamma$ $\approx$ 103 mJ mol$^{-1}$ K$^{-2}$) and the Pauli susceptibility ($\chi$ $\approx$ 4$\times$10$^{-4}$) are strongly enhanced, which confirm the existence of heavy quasiparticles inferred from previous de Haas-van Alphen and ARPES experiments. We discuss this large enhancement using a Gutzwiller slave-boson mean-field calculation, which reveals the proximity of KFe$_{2}$As$_{2}$ to an orbital-selective Mott transition. The temperature dependence of the magnetic susceptibility and the thermal expansion provide  strong experimental evidence for the existence of a coherence-incoherence crossover, similar to what is found in heavy fermion and ruthenate compounds, due to Hund's coupling between orbitals.
\end{abstract}
\maketitle


%
Soon after the discovery of high-temperature superconductivity in iron pnictide compounds,~\cite{Kamihara08,Rotter08} their unique electronic structure, displaying electron and hole sheets, was revealed. In the Ba$_{1-x}$K$_{x}$Fe$_{2}$As$_{2}$ series, superconductivity emerges in the vicinity of an antiferromagnetic spin density wave (SDW) instability (x $\approx$ 0.3) and is maximal at x $\approx$ 0.4 with T$_{c}$ $\approx$ 38 K. At this optimal concentration, the superconducting order parameter is fully gapped with either $s^{++}$ or $s^{+-}$ symmetry.~\cite{Mazin08,Onari09} In the latter case, it is believed that pairing is due to repulsive interband interactions enhanced by the magnetic fluctuations which develop around the nesting vector that connects the two different sheets.~\cite{Mazin08} However, superconductivity is not confined to this hypothetical quantum critical region and persists to x = 1 (with a strongly depressed T$_{c}$ $\approx$ 3 K),~\cite{Rotter08} where only hole pockets are present.~\cite{Terashima10dHvA,Terashima10dHvA2,Yoshida11,Yoshida12} Moreover, the Sommerfeld coefficient for KFe$_{2}$As$_{2}$, $\gamma$ $\approx$ 100 mJ mol$^{-1}$ K$^{-2}$, is paradoxally about twice larger than observed at the optimal concentration~\cite{Fukazawa11,Popovich10} and recent laser ARPES measurements reveal that some of the energy gaps have nodes.~\cite{Okazaki12} Clearly, more experimental investigations are necessary to elucidate the situation in the overdoped region of Ba$_{1-x}$K$_{x}$Fe$_{2}$As$_{2}$ and to understand the origin of the strong mass enhancement observed in quantum oscillation and ARPES experiments in KFe$_{2}$As$_{2}$. LDA+DMFT calculations stress that the mass enhancement in iron pnictides is not related to the proximity to a quantum critical point, but because the electrons, being rather localized at high temperature, start to form coherent quasiparticle bands with the underlying Fermi surface.~\cite{Haule09,Yin11} In this scenario, Hund's rule coupling is responsible for the large mass enhancement and a coherence-incoherence crossover is expected to occur for increasing temperature.  

In this Letter, we combine resistivity, specific-heat, thermal-expansion and susceptibility measurements to study in detail the normal state of KFe$_{2}$As$_{2}$. We clearly show that both the Sommerfeld coefficient ($\gamma$ $\approx$ 102 mJ mol$^{-1}$ K$^{-2}$) and the Pauli susceptibility ($\chi$ $\approx$ 4$\times$10$^{-4}$) are strongly enhanced in this material. Further, our susceptibility and thermal-expansion measurements provide strong evidence for the existence of a crossover between a low-temperature heavy Fermi liquid and a high-temperature incoherent behavior reminiscent of heavy fermion physics. We use a Gutzwiller slave-boson mean-field analysis to obtain a more microscopic understanding of the mass enhancement and our results reveal that KFe$_{2}$As$_{2}$ is indeed close to an orbital-selective Mott transition (OSMT), which qualitatively explains the measured temperature dependences of the susceptibility and thermal expansion.   

\begin{figure}[b]
\begin{center}
\includegraphics[width=8.3cm]{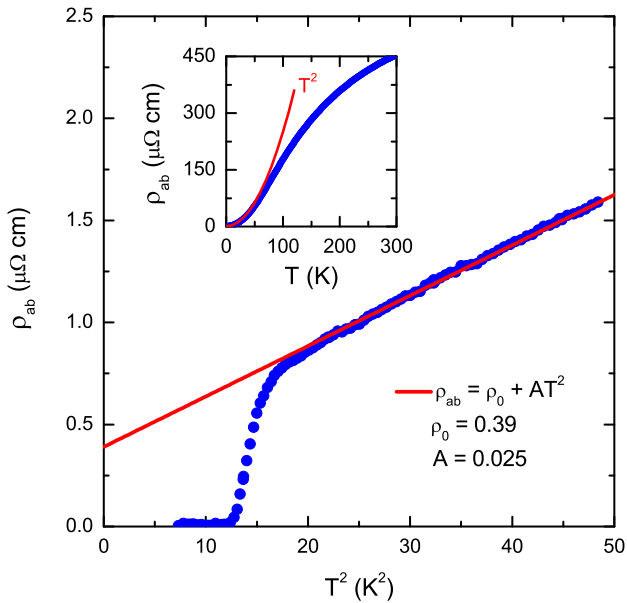}
\caption{\label{fig:Fig1} (Color online) Low-temperature electrical resistivity of KFe$_{2}$As$_{2}$ as a function of T$^{2}$. The solid line is a fit to the usual Fermi liquid behavior, $\rho_{0}$ + AT$^{2}$. Units of $\rho_{0}$ and A are $\mu\Omega$ cm and $\mu\Omega$ cm K$^{-2}$, respectively. The inset shows the temperature dependence of $\rho(T)$ to 300 K.}
\end{center}
\end{figure}

Figure \ref{fig:Fig1} shows the low-temperature electrical resistivity $\rho$(T) of KFe$_{2}$As$_{2}$ measured in the {\it ab} plane. A clear T$^{2}$ dependence, with A $\approx$ 0.025 $\mu\Omega$ cm K$^{-2}$, characteristic of a coherent Fermi liquid is observed. Although the resistive transition appears rather broad ($\Delta$T$_{c}$ $\approx$ 0.5 K), the estimated residual-resistivity ratio, RRR = $\rho$(300K)/$\rho$(0K) $\approx$ 1150, is surprisingly large, in agreement with previous reports.~\cite{Terashima09,Hashimoto10} Such seemingly contradictory behavior suggest that the large RRR results not from an extremely low defect density, but rather from an anomalously high scattering rate at high temperature, as has been predicted by LDA+DMFT calculations with a Hund's coupling constant with {\it e.g.} J$_{H}$ $\approx$ 0.35 eV.~\cite{Haule09,Yin11} We note that a similar situation has already been pointed out for the heavy fermion superconductor URu$_{2}$Si$_{2}$.~\cite{Hassinger11} The resistivity of KFe$_{2}$As$_{2}$ however keeps increasing with temperature in contrast to heavy fermions for which a broad maximum is typically observed around the onset of coherence.\\

\begin{figure}[t]
\begin{center}
\includegraphics[width=8.3cm]{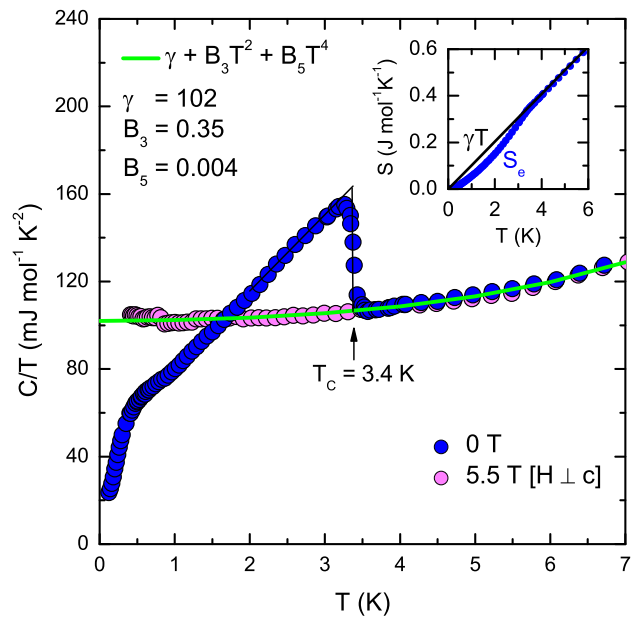}
\caption{\label{fig:Fig2} (Color online) Temperature dependence of the heat capacity of KFe$_{2}$As$_{2}$ in 0 and 5.5 T. The green line represents a fit to the normal-state heat capacity. The Sommerfeld coefficient $\gamma$, the Debye term B$_{3}$ and B$_{5}$ are given in mJ mol$^{-1}$ K$^{-2}$, mJ mol$^{-1}$ K$^{-4}$ and mJ mol$^{-1}$ K$^{-6}$, respectively. The inset shows the normal- and superconducting-state electronic entropies.}
\end{center}
\end{figure}

Figure \ref{fig:Fig2} shows our sample heat capacity down to 0.2 K. A clear jump at T$_{c}$ = 3.4 K, with $\Delta C/ T_{c}$ $\approx$ 54 mJ mol$^{-1}$ K$^{-2}$, marks the transition to the bulk superconducting state. These values are in close agreement with results from Fukazawa \etal\cite{Fukazawa11}. The excess heat capacity observed below 1 K and the modest normalized jump $\Delta C/\gamma T_{c}$ $\approx$ 0.53 with respect to the BCS value (= 1.43) are clear hallmarks of the existence of several energy gaps; the overall curve bears a strong similarity to that of MgB$_{2}$.~\cite{Bouquet01} The solid line in Fig.\ref{fig:Fig2} represents a fit to the normal-state heat capacity from which $\gamma$ = 102 mJ mol$^{-1}$ K$^{-2}$ and the Debye temperature $\Theta_{D}$ $\approx$ 177 K are determined. $\gamma$ is almost 9 times larger than the bare LDA value in KFe$_{2}$As$_{2}$ while only a factor of about 2 is required to reconcile band-structure calculations with calorimetric measurements in Ba(Fe$_{1-x}$Co$_{x}$)$_{2}$As$_{2}$ for 0 $<$ x $<$ 0.4.~\cite{Meingast12} Along with the large Kadowaki-Woods ratio, A/$\gamma^{2}$ $\approx$ 2 $\times$ 10$^{-6}$ $\mu\Omega$ cm K$^{2}$ mol$^{2}$ mJ$^{-2}$, the large $\gamma$ evidently proves the existence of strong correlations in KFe$_{2}$As$_{2}$. Our value agrees well with the existence of heavy bands detected by de Haas-van Alphen and ARPES experiments, which estimate $\gamma$ to be roughly 84 mJ mol$^{-1}$ K$^{-2}$.~\cite{Terashima10dHvA,Terashima10dHvA2,Yoshida11,Yoshida12} Previously such unusually large $\gamma$ values in {\it d} metals have only been found in LiV$_{2}$O$_{4}$ and Sr$_{2}$RuO$_{4}$, for which $\gamma$ reaches 420 and 37.5 mJ mol$^{-1}$ K$^{-2}$, respectively.~\cite{Kondo97,Nishizaki00}

In strongly correlated systems, an enhanced $\gamma$ with respect to the LDA value reflects the importance of correlations. It is often related to a characteristic temperature T$^{*}$ $\propto$ $\gamma^{-1}$ which can be much lower than the bare electronic scale $\epsilon_{f}$/k$_{B}$ because of these correlations. For T $<<$ T$^{*}$, quasiparticles are well defined, and the electronic heat capacity C$_{e}$ is linear in temperature. For T $>>$ T$^{*}$, quasiparticles become short lived: the Landau Fermi-liquid description no longer applies and the heat capacity decreases with temperature to reach ultimately the undressed LDA value. For instance, T$^{*}$ is referred to as the Kondo (or coherence) temperature in heavy fermion systems. Using susceptibility and dilatometry measurements, we show hereafter that this crossover is clearly observed in KFe$_{2}$As$_{2}$. 
    
\begin{figure}[b]
\includegraphics[width=8.3cm]{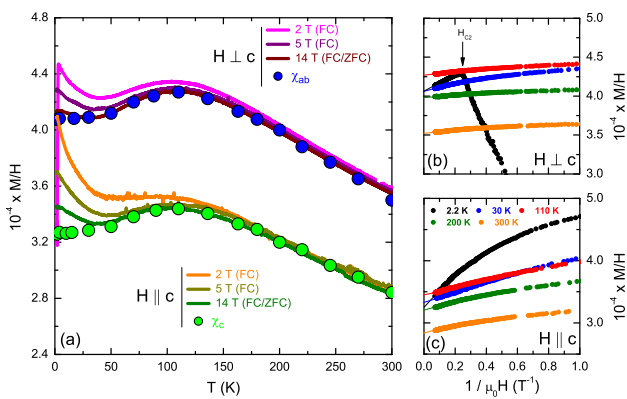}
\caption{\label{fig:Fig4} (Color online) (a) Measured magnetic suceptibility M/H of KFe$_{2}$As$_{2}$ for H $\perp$ c and H $\parallel$ c for several magnetic fields (solid lines). The symbols shows the temperature dependence of the intrinsic suceptibilities $\chi_{c}$(T) and $\chi_{ab}$(T) of KFe$_{2}$As$_{2}$ derived from magnetic isotherms. (b)-(c) M(H)$|_{T}$ isotherms measurements performed at various temperatures, shown in a Honda-Owen representation.}
\end{figure}

Figure \ref{fig:Fig4}(a) shows the temperature dependence of the raw magnetic suceptibility M/H of KFe$_{2}$As$_{2}$ measured for H $\parallel$ c and H $\perp$ c at several constant magnetic fields. For both orientations, M/H increases with decreasing temperatures and exhibits a broad maximum around 100 K. At lower temperatures, a significant upturn, that vanishes with increasing field, develops. This upturn is typical of iron pnictide superconductors and reveals the presence of a small amount of magnetic or paramagnetic impurities, {\it e.g.} Fe atoms. Hence, M(T)$|_{H}$ curves cannot be used to derive the intrinsic susceptibilities $\chi_{ab}$(T) and $\chi_{c}$(T) of KFe$_{2}$As$_{2}$. Alternatively, they can be determined precisely by analyzing magnetic isotherms, M(H)$|_{T}$, performed at various temperatures. As argued by Johnston,~\cite{Johnston10} one has in this case:
\begin{equation}\label{eq:eq2}
M(H)|_{T}=\chi(T)H+M_{i}(H)|_{T},
\end{equation}
where $\chi$(T) and M$_{i}(H)|_{T}$ are the intrinsic susceptibility of KFe$_{2}$As$_{2}$ and the impurity contribution to M(H)$|_{T}$, respectively. Since  M$_{i}(H)|_{T}$ is not linear with field and typically saturates at some field, $\chi$(T) represents the slope of the high-field linear region of the measured M(H)$|_{T}$ curves. Figures \ref{fig:Fig4}(b) and \ref{fig:Fig4}(c) show several M(H)$|_{T}$ isotherms for both orientations displayed in a Honda-Owen representation~\cite{Honda10,Owen12} which gives the intrinsic $\chi$(T) by extrapolating M/H to 1/H $\rightarrow$ 0, according to Eq.\ref{eq:eq2}. The derived intrinsic susceptibilities for H $\parallel$ c and H $\perp$ c are shown in Fig.\ref{fig:Fig4}(a) as solid circles.
For T $<$ 50 K, both $\chi_{c}$ and $\chi_{ab}$ are nearly temperature independent and rather large, equal to 3.2$\times$10$^{-4}$ and 4.1$\times$10$^{-4}$, respectively. This Pauli paramagnetic behavior is consistent with the Fermi liquid behavior observed in resistivity and heat-capacity measurements. With these values, we find a Wilson ratio R$_{W}=\frac{\pi^{2}k_{B}^{2}V_{m}\chi}{3\mu_{0}\mu_{B}^{2}\gamma_{e}}$ of 1.1 and 1.3 for H $\parallel$ c and H $\perp$ c, respectively. A Wilson ratio close to unity indicates that the same quasiparticles are involved in the enhancement of both $\gamma$ and the Pauli susceptibility. Moreover, the magnetic susceptibility is weakly anisotropic with a temperature-independent ratio $\chi_{ab}$/$\chi_{c}$ $\approx$ 1.3, which is in good agreement with the Knight shift ratio, K$_{ab}$/K$_{c}$ $\approx$ 1.2 - 1.5, extracted from NMR measurements.~\cite{Zhang10,Hirano12} Finally, for T $>$ 50 K, $\chi_{ab}$ and $\chi_{c}$ increase and reach a broad maximum around 100 K and then monotonically decrease beyond, following approximately a Curie-Weiss law, $\chi$=$\frac{C}{T+\Theta_{cw}}$, for T $>$ 150 K from which a fluctuating paramagnetic moment of about 2.5 $\mu_{B}$ is inferred and $\Theta_{cw}$ $\approx$ 600 K which provides a crude estimation of T$^{*}$. This overall behavior is extremely similar to that of the heavy fermion CeRu$_{2}$Si$_{2}$ which is paramagnetic but close to an antiferromagnetic instability.~\cite{Haen87}   

\begin{figure}[t] 
\begin{center}
\includegraphics[width=8.3cm]{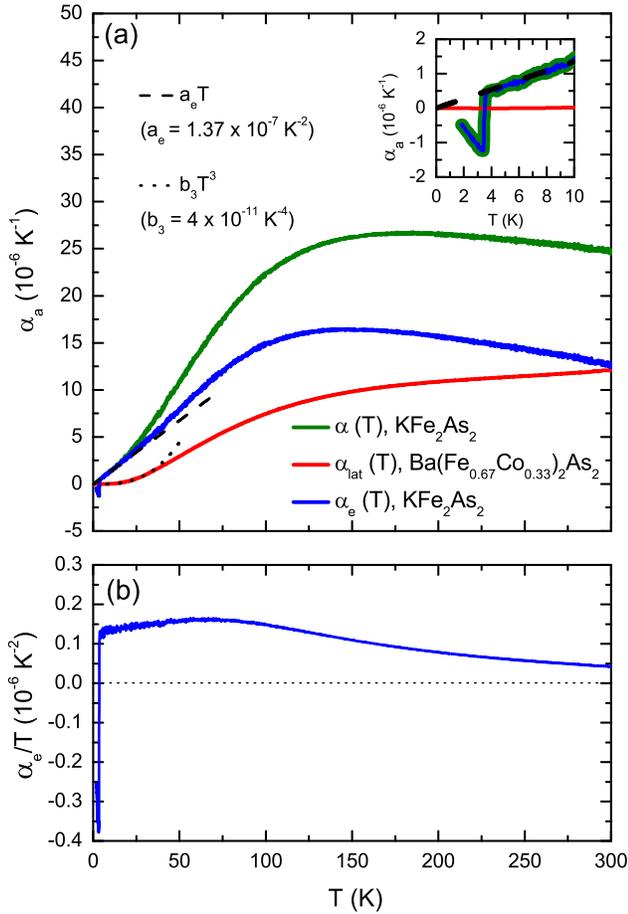}
\caption{\label{fig:Fig3} (Color online) (a) Temperature dependence of the thermal expansion of KFe$_{2}$As$_{2}$ and Ba(Fe$_{0.67}$Co$_{0.33}$)$_{2}$As$_{2}$ measured along the {\it a} axis. The dashed and dotted lines are low-temperature fits to the Fermi liquid ($a$T) and Debye (b$_{3}$T$^{3}$) models, respectively. The blue line shows the electron thermal expansion of KFe$_{2}$As$_{2}$ obtained by subtracting the expansivity of Ba(Fe$_{0.67}$Co$_{0.33}$)$_{2}$As$_{2}$ from that of KFe$_{2}$As$_{2}$. The inset shows a magnified view of the low-temperature regime. (b) Temperature dependence of $\alpha_{e}$/T. }
\end{center}
\end{figure}

In CeRu$_{2}$Si$_{2}$, the electronic heat capacity C$_{e}$/T is constant at low temperature with an enhanced $\gamma$ = 360 mJ mol$^{-1}$ K$^{-2}$ and then decreases monotonically for T $>$ 5 K.~\cite{Fisher88} Thus, a similar behavior should be observed for KFe$_{2}$As$_{2}$ but this crossover is complicated to address experimentally in calorimetric measurements because of the large phonon background. On the other hand, measurements of the coefficient of linear thermal expansion, 
\begin{equation}\label{eq:eq1}
\alpha_{i}(T)=\frac{1}{L_{i}}\left(\frac{\partial L_{i}}{\partial T}\right)_{p_{i}}=-\left(\frac{\partial S}{\partial p_{i}}\right)_{T} (i=a,c),
\end{equation}
which probe the changes of entropy with respect to uniaxial pressure $p_{i}$, often show a relatively larger electronic signal compared to the heat capacity. This is particularly true for the iron-pnictide superconductors, which show large electronic signatures in the thermal expansion.~\cite{Meingast12,Hardy09} Figure \ref{fig:Fig3} shows the in-plane thermal expansion of KFe$_{2}$As$_{2}$ and, at low temperature, a clear Fermi-liquid behavior $\alpha$ = $a$T, with a large electronic term a = 1.37 $\times$ 10$^{-7}$ K$^{-2}$ is observed. The raw data also clearly demonstrate that the low-temperature linear behavior does not survive up to higher temperatures, since the data near room temperature are actually decreasing with increasing temperature. In order to get a more quantitative measure of the electronic signal, we subtracted the thermal expansion of Ba(Fe$_{0.67}$Co$_{0.33}$)$_{2}$As$_{2}$, which previously was shown to have almost no electronic term and is thus a good approximation of the thermal expansion due to anharmonic phonons.~\cite{Meingast12} The resulting electronic term, $\alpha_{e}$(T), increases linearly at low temperatures, passes over a maximum near 125 K and then further decreases up to 300 K. This behavior clearly demonstrates the expected crossover from a low- temperature Fermi liquid regime to a high-temperature regime with a strongly reduced electronic thermal expansion. In Fig.\ref{fig:Fig3}(b), we show $\alpha_{e}$/T, whose temperature dependence bears a strong similarity to that of the susceptibilities shown in Fig.\ref{fig:Fig4}(a). Assuming C$_{e}$ $\propto$ $\alpha_{e}$, Fig.\ref{fig:Fig3}(b) also reflects the temperature variation of the electronic heat capacity C$_{e}$/T. Thus, our thermal-expansion and susceptibility measurements both show clear evidence of the coherence-incoherence crossover predicted by LDA+DMFT calculations.~\cite{Haule09} 
Although there are clear similarities between KFe$_{2}$As$_{2}$ and heavy fermions, the direct comparison is probably too simplistic since there are no 4$\it f$ electrons involved in this iron pnictide. Nevertheless, our measurements clearly illustrate the existence of strong antiferromagnetic correlations in KFe$_{2}$As$_{2}$. Beside large $\gamma$ values, similar temperature dependences were reported in other correlated {\it d}-metal alloys, {\it e.g.} the layer perovskite Ca$_{2-x}$Sr$_{x}$RuO$_{4}$.~\cite{Nakatsuji03} In these materials, the origin of the heavy-fermion like behavior is still unsettled. But theoretical calculations (DMFT+LDA) ascribe it to an orbital-selective Mott transition (OSMT) since hybridization induces orbital fluctuations that results in the formation of a Kondo-like heavy-fermion behavior.~\cite{Koga05} The proximity to such an OSMT is actually debated in iron pnictide superconductors.

\begin{figure}[b] 
\begin{center}
\includegraphics[width=8.3cm]{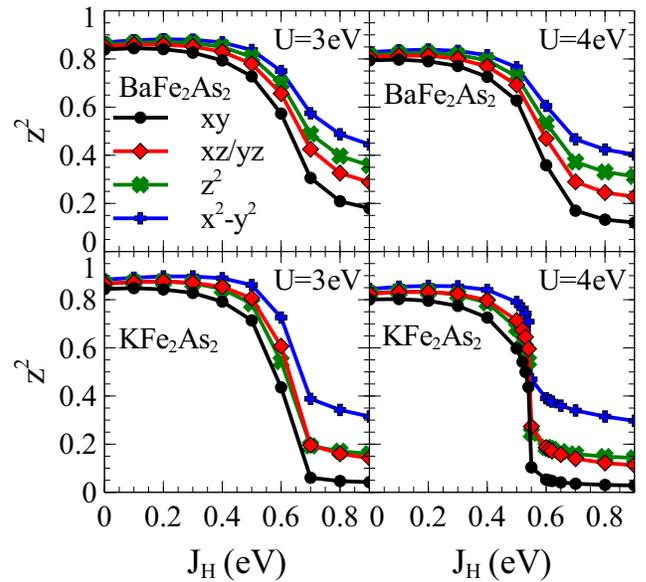}
\caption{\label{fig:Fig5} (Color online) Gutzwiller slave-boson mean-field calculations of the orbitally-resolved mass-enhancement factor $z_{\alpha}$ for BaFe$_{2}$As$_{2}$ and KFe$_{2}$As$_{2}$, for two different values of the intraorbital Coulomb repulsion $U$.}
\end{center}
\end{figure}

In order to obtain a more microscopic understanding of the origin of the significant mass enhancement in KFe$_{2}$As$_{2}$, we performed a Gutzwiller corrected electronic structure calculation following our earlier analysis for related systems.~\cite{Yao11} Using a tight-binding fit of the iron {\it3d} bands, we include local intra- and interorbital Coulomb repulsions $U$ and $U'$, respectively, as well as Hund's coupling J$_{H}$ between the various electrons. In order to describe and characterize electronic correlation effects we performed a Gutzwiller slave-boson mean-field calculation that yields orbitally resolved mass enhancement factors z$_{\alpha}$ where $\alpha$ refers to the five {\it d}-orbitals. Our results are shown in Fig.\ref{fig:Fig5} where we plot z$_{\alpha}$(J$_{H}$) for different values of {\it U} for KFe$_{2}$As$_{2}$ and BaFe$_{2}$As$_{2}$.
In both materials, the Fermi surface sheets of $\left|xy\right\rangle$ character are the most strongly coupled, followed by $\left|xz\right\rangle$, $\left|yz\right\rangle$ and $\left|z^{2}\right\rangle$ states. These overall trends are similar to what is found in LDA-DMFT calculations.~\cite{Yin11} One important distinction between BaFe$_{2}$As$_{2}$ and KFe$_{2}$As$_{2}$ is clearly the more prominent role played by $\left|xy\right\rangle$ states on the Fermi surface. We also find a significant difference in the values of the effective masses $m^{*}$ $\propto$ $z_{\alpha}^{-1}$ for KFe$_{2}$As$_{2}$ in comparison with BaFe$_{2}$As$_{2}$ if we consider sufficiently large values of the local Coulomb interaction and the Hund's coupling. For U $\gtrsim$ 4 eV we find for J$_{H}$ $\gtrsim$ 0.5-0.6 eV an orbitally selective Mott transition with a localization of the $\left|xy\right\rangle$ states. While these parameters are too large for a realistic description of the material, they demonstrate the proximity of KFe$_{2}$As$_{2}$ to an OSMT for the configuration intermediate between $3d^{5}$ and $3d^{6}$, a concept that was already discussed in the pnictide family in Refs~\onlinecite{Yao11} and~\onlinecite{Misawa12}. The fact that the significant mass enhancement occurs for $\left|xy\right\rangle$ dominated bands is in complete agreement with quantum oscillation~\cite{Terashima10dHvA,Terashima10dHvA2} and ARPES experiments~\cite{Yoshida11,Yoshida12}. A well known situation that can be rationalized as an OSMT is the heavy fermion behavior in Kondo alloys. Similarly to those materials it is rather natural to expect local moment behavior at high temperatures and Fermi liquid behavior with enhanced mass at low temperatures in the present situation. Thus, one would generally expect a temperature dependence of the resistivity, susceptibility and heat capacity in KFe$_{2}$As$_{2}$ that closely resembles that of a heavy fermion systems.

In summary, by studying the normal-state properties of KFe$_{2}$As$_{2}$, we have shown that this is a strongly correlated material with highly renormalized values of both the Sommerfeld coefficient and the Pauli susceptibility. Additionally, we have provided evidence for a temperature-induced incoherent-coherent crossover to this highly correlated state. The physics of these correlations and this crossover are well described by the possible close proximity of KFe$_{2}$As$_{2}$ to an orbital-selective Mott transition due to Hund's coupling between orbitals. It is interesting to note that these strong correlations do not enhance superconductivity - T$_{c}$ is rather low and does not correlate with $\gamma$ for different K dopings. In fact, the correlations appear to be detrimental to superconductivity, which may be a rather general feature of high-temperature superconductivity, {\it e.g.} in the 115 class of compounds, the highest T$_{c}$ $\approx$ 18 K is found for the weakly correlated Pu compound~\cite{Bauer04} while CeIrIn$_{5}$, with a large $\gamma$ $\approx$ 800 mJ mol$^{-1}$ K$^{-2}$, shows a quite low T$_{c}$.~\cite{Movshovich11}

\begin{acknowledgments}
We thank J. Flouquet, K.- M. Ho and X. Z. Wang for stimulating and enlightening discussions. This work was supported by the Deutsche Forschungsgemeinschaft through DFG-SPP 1458 "Hochtemperatursupraleitung in Eisenpniktiden". Y. X. Yao gratefully acknowledges the support of U.S. Department of Energy through the Computational Materials and Chemical Sciences Network. The work in Grenoble (D. Aoki) was supported by the ERC starting grant "NewHeavyFermion".    
\end{acknowledgments}

\bibliography{biblio}
\end{document}